%% file: Main.tex
\documentclass[sigconf]{utility/acmart}

\input{utility/preamble}
\input{utility/metadata}

\AtBeginDocument{}

\makeatletter 
\ifcameraready\else 
\def\ACM@cc@type{} 
\renewcommand{\@copyrightpermission}{} 
\renewcommand{\@mkbibcitation}{ 
    \par\medskip\small\noindent{\bfseries ACM Reference Format:}\par\nobreak \noindent\bgroup \def\\{\unskip{}, \ignorespaces}\authors\egroup. \@acmYear. \@title. In \textit{\@acmBooktitle}, \ref{TotPages}~pages. \ifx\@acmDOI\@empty\else\space\@formatdoi{\@acmDOI}\fi 
    \par } 
\fi 
\makeatother

\begin{document}
\title{Knowledge Graph Construction for Stock Markets \\
with LLM-Based Explainable Reasoning}
\input{sections/0.Authors}

\renewcommand{\shortauthors}{Cheonsol Lee et al.}

\input{sections/0.Abstract}

\input{sections/0.CCS_Keyword}

\maketitle

\input{sections/1.Introduction}
\input{sections/2.Related_Works}
\input{sections/3.Methodology}
\input{sections/4.Case_Study}
\input{sections/5.Discussion_Conclusion}


\bibliographystyle{utility/ACM-Reference-Format}
\bibliography{sections/Reference}

\appendix
\input{sections/6.Appendix}





\end{document}

%% file: utility/preamble.tex
\usepackage{booktabs}
\usepackage{microtype}
\usepackage{graphicx}
\usepackage{enumitem}
\usepackage{afterpage}
\usepackage{listings}
\usepackage{xcolor}
\usepackage{minted}
\graphicspath{{figs/}}
\usepackage{caption}
\usepackage{longtable}
\usepackage{array}
\lstdefinestyle{sqlstyle}{
    language=SQL,
    basicstyle=\ttfamily\small,
    keywordstyle=\color{blue},
    stringstyle=\color{red},
    commentstyle=\color{green},
    frame=single,
    breaklines=true,
    numbers=left,
    escapeinside={(*@}{@*)} 
}

\renewcommand{\shortauthors}{Cheonsol Lee et al.}


%% file: utility/metadata.tex
\newif\ifcameraready     

\newif\ifanonymous        

\setcopyright{acmlicensed}

\copyrightyear{2025}
\acmYear{2025}
\acmDOI{}   
\acmISBN{}  

\newcommand{\ConfAcronym}{CIKM '25}
\newcommand{\ConfFull}{Proceedings of the 34th ACM International Conference on Information and Knowledge Management}
\newcommand{\ConfDates}{November 10--14, 2025}
\newcommand{\ConfLoc}{Seoul, Republic of Korea}

\acmConference[\ConfAcronym]{\ConfFull}{\ConfDates}{\ConfLoc}

\acmBooktitle{\ConfFull\ (\ConfAcronym), \ConfDates, \ConfLoc}

\ifcameraready
\else
\fi

\ifanonymous
\fi


%% file: sections/0.Authors.tex
\author{Cheonsol Lee}
\orcid{0000-0001-7169-0035}
\affiliation{%
    \institution{Hana Institute of Technology, Hana TI}
    \city{Seoul}
    \country{Republic of Korea}
}
\email{cheonsol.lee@hanafn.com}

\author{Youngsang Jeong}
\affiliation{%
    \institution{TelePIX Co., Ltd.}
    \city{Seoul}
    \country{Republic of Korea}
}
\email{videorighter@gmail.com}

\author{Jeongyeol Shin}
\affiliation{%
    \institution{Qraft Technologies}
    \city{Seoul}
    \country{Republic of Korea}
}
\email{sinjy1203@gmail.com}

\author{Huiju Kim}
\affiliation{%
    \institution{BISTelligence}
    \city{Seoul}
    \country{Republic of Korea}
}
\email{wewe828@gmail.com}

\author{Jidong Kim}
\affiliation{%
    \institution{TelePIX Co., Ltd.}
    \city{Seoul}
    \country{Republic of Korea}
}
\email{apitots@telepix.net}

%% file: sections/0.Abstract.tex
\begin{abstract}

The stock market is inherently complex, with interdependent relationships among companies, sectors, and financial indicators. Traditional research has largely focused on time-series forecasting and single-company analysis, relying on numerical data for stock price prediction. While such approaches can provide short-term insights, they are limited in capturing relational patterns, competitive dynamics, and explainable investment reasoning. To address these limitations, we propose a knowledge graph schema specifically designed for the stock market, modeling companies, sectors, stock indicators, financial statements, and inter-company relationships. By integrating this schema with large language models (LLMs), our approach enables multi-hop reasoning and relational queries, producing explainable and in-depth answers to complex financial questions. Figure\,\ref{fig:pipeline} illustrates the system pipeline, detailing the flow from data collection and graph construction to LLM-based query processing and answer generation. We validate the proposed framework through practical case studies on Korean listed companies, demonstrating its capability to extract insights that are difficult or impossible to obtain from traditional database queries alone. The results highlight the potential of combining knowledge graphs with LLMs for advanced investment analysis and decision support.

\end{abstract}

%% file: sections/0.CCS_Keyword.tex
\ccsdesc[500]{Computing methodologies~Knowledge representation and reasoning}
\ccsdesc[500]{Computing methodologies~Semantic networks}
\ccsdesc[500]{Computing methodologies~Ontology engineering}
\ccsdesc[500]{Information systems~Recommender systems}

\keywords{Knowledge Graph, Financial AI, Recommender Systems, Large Language Models, Conversational Agents, Chatbots, Information Retrieval, Natural Language Processing, Stock Market}

%% file: sections/1.Introduction.tex
\vspace{-0.5cm}

\begin{figure}[H]
  \centering
  \includegraphics[width=0.47\textwidth]{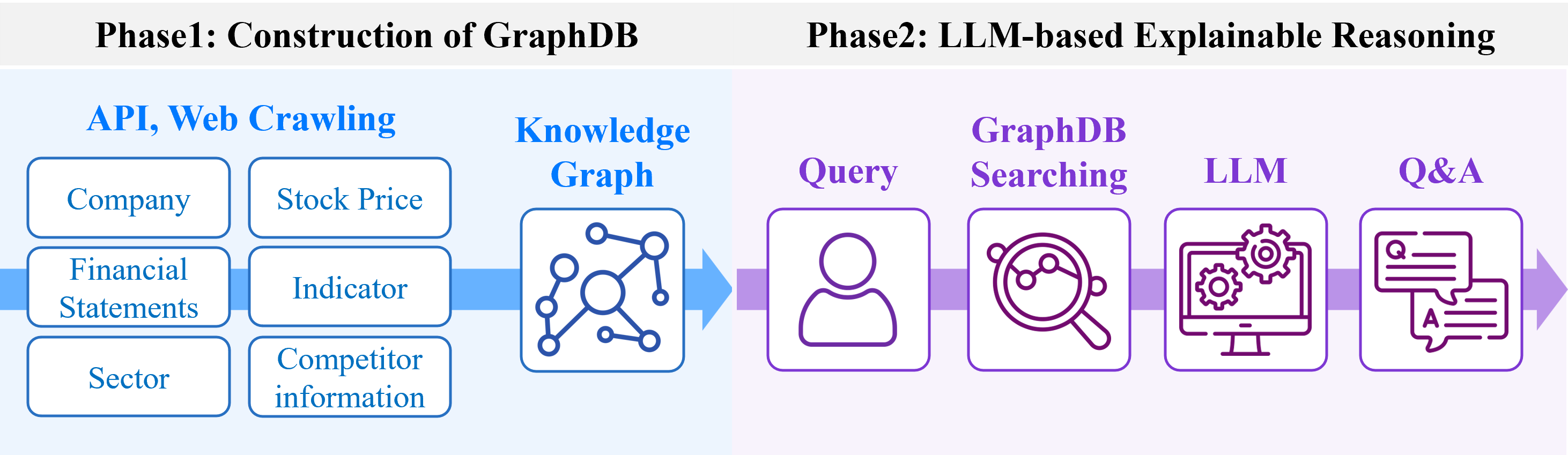}
  \vspace{-0.2cm}
  \caption{A pipeline for GraphDB construction and LLM-based inference system}
  \label{fig:pipeline}
\end{figure}

\section{Introduction}\label{sec1}
Large language models (LLMs) have recently demonstrated remarkable capabilities across various domains, including finance, particularly in question \& answering (Q\&A) and recommender systems. Despite their potential, LLMs are prone to hallucinations, which limits their ability to provide reliable and accurate investment insights. In parallel, knowledge graphs have been widely recognized for their ability to represent complex relationships and support explainable reasoning. Graph Retrieval Augmented Generation (GraphRAG) combines these two approaches, leveraging LLMs to query relational data in knowledge graphs and generate interpretable insights. This capability is particularly valuable for financial applications, where understanding inter-company relationships and market trends is as important as predicting stock prices.

Existing research in the financial domain has largely focused on stock price prediction using time-series models such as ARIMA, LSTM, Transformer-based architectures \cite{ariyo2014stock, selvin2017stock, muhammad2023transformer}, and graph neural networks (GNNs) to capture stock correlations \cite{shi2024integrated}. LLMs have also been explored for predicting stock trends and portfolio optimization \cite{wawer2025integrating, xu2025modeling}. While these studies achieve impressive predictive performance, they predominantly focus on numerical accuracy and provide limited interpretability, often overlooking relational structures between companies and sectors.

However, despite the promise of combining LLMs with relational data, the application of graph databases (GraphDBs) in the stock domain remains limited. Constructing a comprehensive stock market knowledge graph requires significant effort to integrate company attributes, financial statements, stock indicators, and inter-company relationships. As a result, research on GraphDB-based recommender or Q\&A systems for stock analysis is still in its infancy, especially for the Korean market.

\begin{figure}[H]
    \centering
    \includegraphics[width=0.47\textwidth]{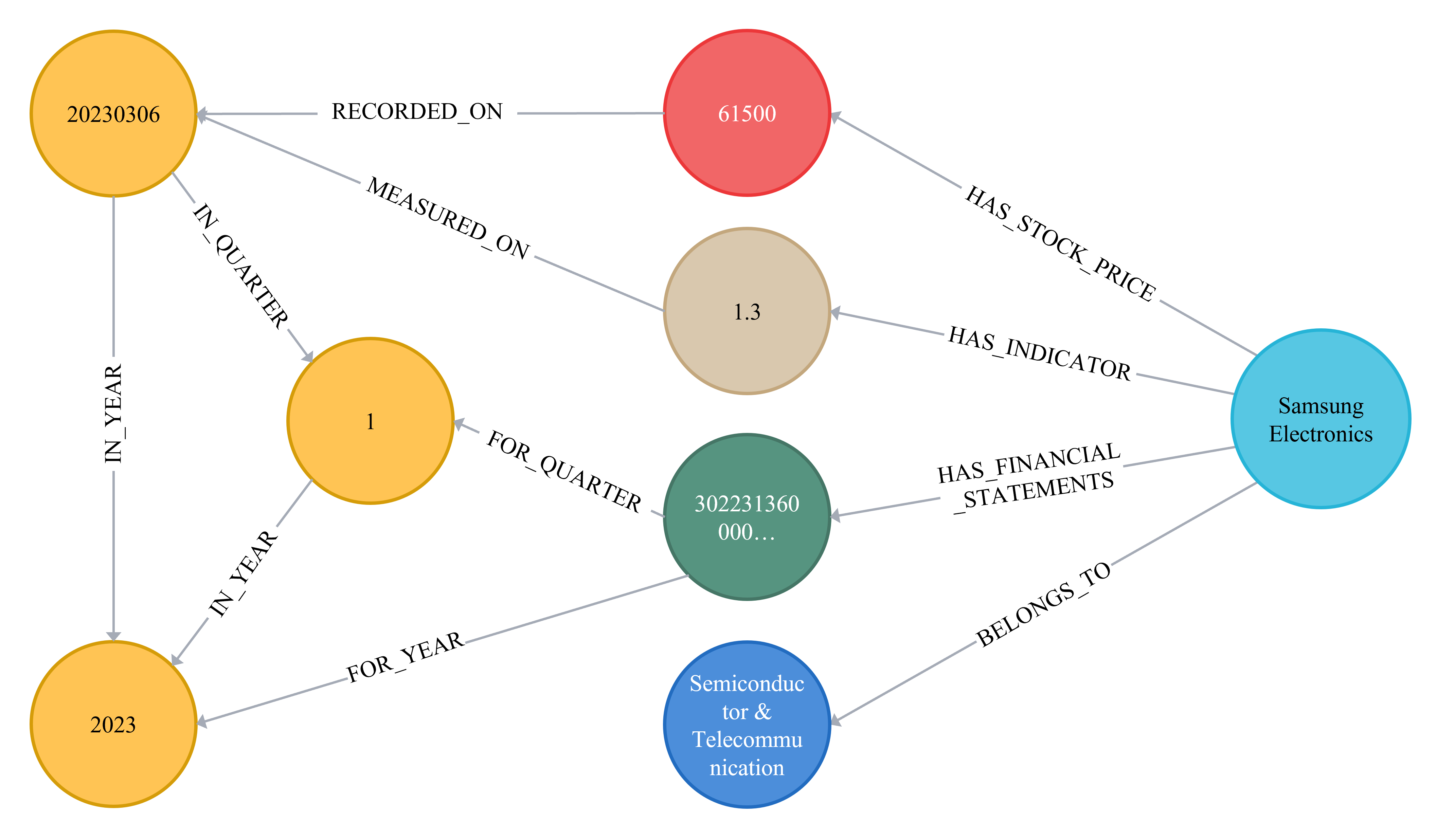}
    \vspace{-0.2cm}
    \caption{An example of the graph database}
    \Description{Graph example illustration}
    \label{fig:graph_example}
\end{figure}

To address this gap, we propose a knowledge graph schema specifically designed for the stock market, a novel framework that builds a multi-relational GraphDB using data from Korean listed companies and integrates it with LLMs for explainable reasoning. Our knowledge graph represents companies, industries, stock prices, financial indicators, and financial statements as nodes and edges, enabling multi-hop queries that reveal hidden patterns and interdependencies. For instance, a natural language query regarding Samsung Electronics (stock code: 005930) is automatically transformed into a Cypher query to retrieve relevant information from the Neo4j server, as illustrated in Listing\,\ref{lst:cypher-query}. Figure\,\ref{fig:graph_example} depicts a sample graph enabling relational inquiries such as comparative financial analysis and trend evaluation.

\vspace{-0.3cm}

\begin{listing}[H]
    \caption{Cypher query for Samsung Electronics}
    \label{lst:cypher-query}
    \begin{lstlisting}[
      frame=single,
      numbers=left,
      basicstyle=\ttfamily\small,
      columns=fullflexible
    ]
    MATCH (c:Company{stock_code:"005930"})-[:HAS_STOCK_PRICE]
    ->(sp:StockPrice)-[:RECORDED_ON]->(d:Date{date:"20230306"})
    OPTIONAL MATCH (d)-[:IN_YEAR]->(y:Year)
    OPTIONAL MATCH (d)-[:IN_QUARTER]->(q:Quarter)
    OPTIONAL MATCH (c)-[r]-(connected)
    RETURN c, r, connected, sp, d, y, q
    \end{lstlisting}
    \vspace{-0.3cm}
\end{listing}

The main contributions of this study are summarized as follows:
\begin{itemize}[noitemsep, topsep=1pt]
\item \textbf{Framework:} Construction of a comprehensive stock-domain Graph Database integrating Korean stock market data.
\item \textbf{Methodology:} Integration of LLMs with GraphDB for multi-hop, explainable reasoning over financial data.
\item \textbf{Case Study:} Practical stock market case studies demonstrating insights beyond conventional database queries.
\end{itemize}

\vspace{0.3cm}

%% file: sections/2.Related_Works.tex
\section{Related Work}

\noindent\textbf{Research Trends in the Stock Domain}

Research in the stock domain has traditionally focused on time-series forecasting. Classical models such as ARIMA, as well as deep learning approaches including LSTM and Transformer-based models, have been widely applied for stock price prediction \cite{ariyo2014stock, Cheng2024}. Recently, graph neural networks (GNNs) have been proposed to capture correlations between stocks and improve prediction performance \cite{shi2024integrated}. In addition, large language models (LLMs) have been increasingly utilized in financial research, demonstrating effectiveness in both time-series prediction and portfolio optimization tasks \cite{wawer2025integrating, xu2025modeling}. However, these approaches primarily focus on numerical prediction accuracy and are limited in providing explainable insights or relational analyses between companies.

\noindent\textbf{Applications of Graph Databases in the Financial Industry}

Graph databases offer an intuitive representation of complex relationships, enabling analysis that is difficult with traditional relational databases. They have been applied in various domains including social network analysis \cite{he2020constructing}, recommender systems \cite{guo2020survey}, and biological data modeling \cite{xu2020building}. In the financial sector, GraphDB has been employed for fraud detection using transaction networks \cite{zhan2018loan} and for analyzing inter-institutional relationships \cite{zehra2021financial}. While these studies primarily focus on anomaly detection or network structure analysis, few have explored constructing comprehensive stock market knowledge graphs for generating actionable investment insights.

\noindent\textbf{Recommender Systems with Knowledge Graphs}

Knowledge graphs are frequently used in recommender systems due to their ability to represent rich relational information between items. Recent work has focused on providing explainable recommendations by modeling item attributes and user-item interactions as graph structures. Nevertheless, applications of knowledge graphs in financial domains, especially for stock recommendation, remain limited. In particular, no prior studies have addressed knowledge graph-based recommendation for the Korean stock market.

\noindent\textbf{Large Language Models for Graph Reasoning}

LLMs have recently been applied beyond text generation to perform relational reasoning over knowledge graphs. Approaches have been proposed to translate natural language queries into graph query languages such as SPARQL or Cypher, allowing GraphDB to return accurate and explainable answers \cite{yang2026enhancing}. Despite these advances, no prior work has applied this LLM-to-Graph reasoning paradigm to financial data, particularly stock market data.

\noindent\textbf{Summary and Gap}

In this study, we construct a comprehensive GraphDB using Korean stock market data and integrate it with LLMs to enable explainable multi-hop reasoning. Unlike previous studies that focus on either predictive modeling or network anomaly detection, our framework allows relational analysis across companies and financial indicators, providing actionable insights for investors and analysts.

%% file: sections/3.Methodology.tex
\begin{figure*}[t]
    \centering
    \includegraphics[width=\textwidth]{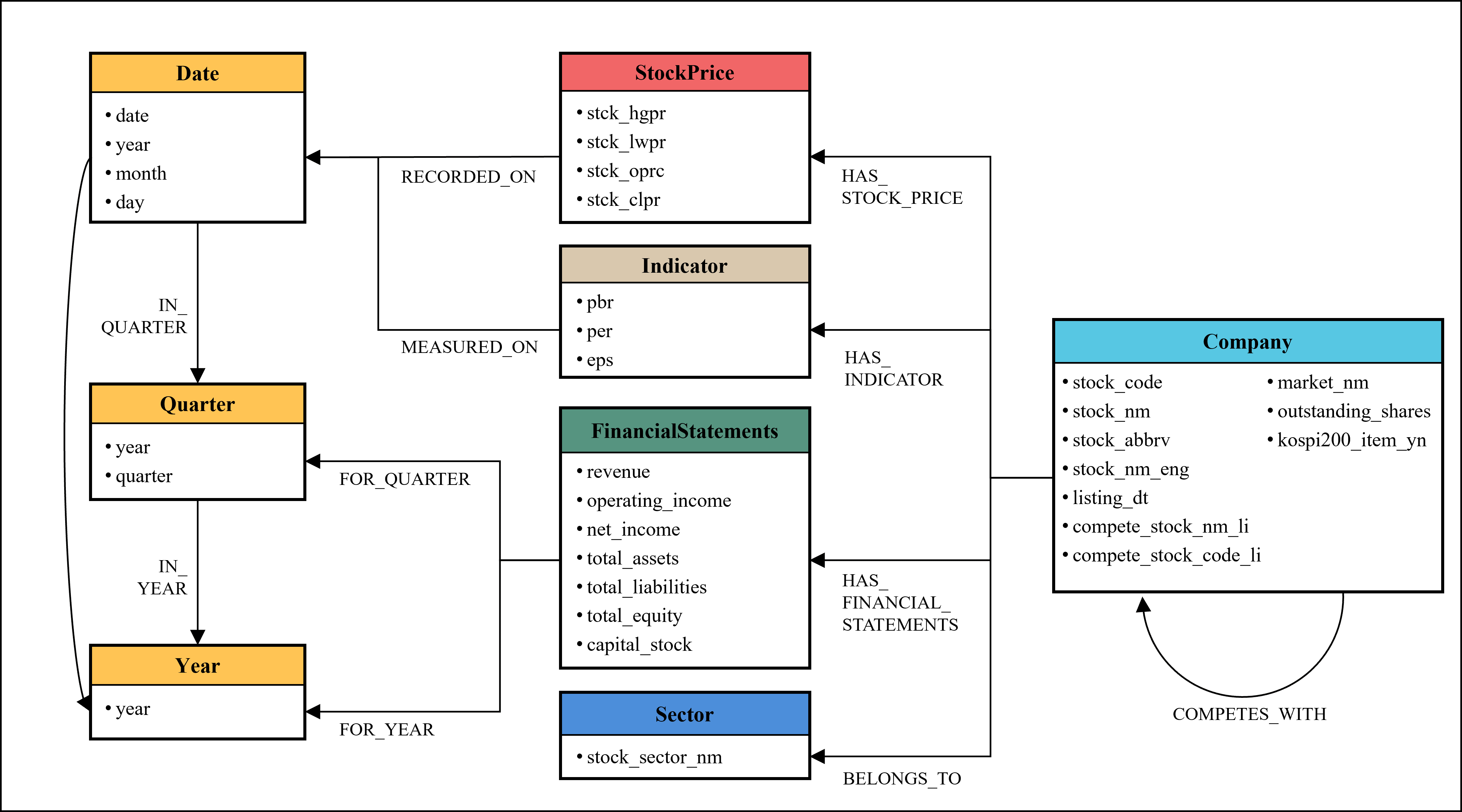}
    \caption{An overview of the proposed graph schema}
    \label{fig:graph_schema}
\end{figure*}

\section{Methodology}\label{sec3}

A pipeline for GraphDB construction with the LLM-based reasoning system is shown in Figure\,\ref{fig:pipeline}. Phase 1 is a process of constructing GraphDB and is described in detail in 3.1, 3.2, and 3.3. Phase 2 is a process of inferring using LLM and is described in 3.4.

\subsection{Data Collection}

This study constructed a knowledge graph of the Korean stock market using publicly available API-based data. Daily stock prices and key investment indicators (e.g., PER, PBR) for individual stocks were collected via the KIS Developers API\footnote{\href{https://apiportal.koreainvestment.com/intro}{Korea
 Investment \& Securities' Open API}}. Financial statement data, including assets, liabilities, revenue, and operating profit, were obtained from the OpenDART API\footnote{\href{https://opendart.fss.or.kr/}{OpenDart
 API}}. Basic company information, such as name, stock code, and industry sector, was retrieved from the Korea Exchange (KRX) API\footnote{\href{https://data.krx.co.kr/}{KRX
 API}}. Competitor relationships were collected from the `Competitor Analysis' section of Company Wise reports\footnote{\href{https://comp.wisereport.co.kr/}{Company
 Wise Report}} and supplemented with web crawling.

The collected data were consolidated by company after pre-processing, resulting in financial data for 2,879 listed companies over a three-year period (2023–2025).

\subsection{Graph Schema Design}

Figure\,\ref{fig:graph_schema} illustrates an overview of the proposed graph schema. The GraphDB schema is represented as a class diagram to facilitate the visualization of nodes, relationships, and their properties, where each object represents a node and each edge represents a relationship between nodes. Table\,\ref{tab:graph_schema} summarizes the types of nodes and relationships included in the schema.

\begin{table}
\centering
\begin{tabular}{|l|l|}
\hline
\textbf{Node type} & \textbf{Relationship type} \\
\hline
Company & HAS\_STOCK\_PRICE \\
Sector & HAS\_INDICATOR \\
Indicator & HAS\_FINANCIAL\_STATEMENTS \\
Stock Price & BELONGS\_TO \\
Financial Statements & COMPETES\_WITH \\
Date & RECORDED\_ON \\
Quarter & MEASURED\_ON \\
Year & FOR\_QUARTER \\
 & FOR\_YEAR \\
 & IN\_YEAR \\
 & IN\_QUARTER \\
\hline
\end{tabular}
\caption{Node types and Relationship types}
\label{tab:graph_schema}
\end{table}

Detailed descriptions of the properties of each node are provided in Table\,\ref{tab:graphdb_property} in the appendix. The names of nodes, relationships, and properties are intuitively labeled, enabling the construction of the knowledge graph.

\subsection{GraphDB Implementation}

The graph database was implemented using Neo4j. Data insertion and relational queries were performed using Neo4j's Cypher query language. The database contains millions of nodes and relationships, including approximately 2,879 publicly traded companies with financial data spanning 2023 to 2025.

\subsection{LLM Integration}

We employed the GPT-4.1 model as the backbone LLM. The system prompt provided the GraphDB schema to the model, while the user prompt included additional rules for query formulation based on the type of question. These rules are detailed in Appendix B. The operation of the LLM consists of the following three steps:

\vspace{0.1cm}
\begin{enumerate}[noitemsep, topsep=1pt]
\item \textbf{Question-to-Cypher Conversion}: In the first step, the input natural language question is converted into a Cypher query using LangChain’s LLMChain with a dedicated prompt. The LLM outputs the corresponding Cypher query.
\item \textbf{Graph Query Execution}: In the second step, the generated Cypher query is executed on the GraphDB, which returns the query results.
\item \textbf{Answer Generation}: In the third step, the original question, the generated Cypher query, and the query results are provided to the LLM along with a specially designed prompt to generate the final answer.
\end{enumerate}

%% file: sections/4.Case_Study.tex
\section{Case Study}\label{sec4}

The case studies presented in this research highlight the practical value and theoretical implications of applying a GraphDB framework, combined with LLM-based query mechanisms, to the analysis of the stock market. Each case not only demonstrates the feasibility of graph-driven financial exploration but also offers distinct contributions to investment analysis and decision-making. For detailed steps of LLM execution and answer generation in the case studies, refer to Appendix~B.

\vspace{0.2cm}
\noindent\textbf{Case Study 1: Multi-year financial comparison between competitors}

The objective of this case study is to analyze the multi-year financial performance of a company and its main competitor within the same industry. Samsung Electronics (stock code: 005930) and SK Hynix (stock code: 000660) were selected for the analysis. Using the GraphDB schema, we traverse the COMPETES\_WITH relationship to identify competitor companies and retrieve the associated FinancialStatements nodes for each company for the years 2023, 2024, and 2025. The key financial metrics considered include revenue, operating income, and net income, which were examined to identify trends and relative performance over the three-year period.

The findings reveal distinct performance patterns for the two companies. Samsung Electronics demonstrates a strong rebound in 2024, with significant increases in both revenue and profitability compared to 2023, followed by a slight decline in 2025. In contrast, SK Hynix exhibits steady growth from 2023 onwards, recovering from initial lower earnings and achieving continuous improvements in revenue, operating income, and net income. By 2025, SK Hynix surpasses Samsung Electronics in operating income and net income, despite Samsung Electronics maintaining a larger overall revenue scale.

This competitive analysis highlights that while Samsung Electronics retains strength in total revenue, SK Hynix shows a pronounced recovery in profitability. GraphDB enables these insights by supporting multi-hop queries that retrieve financial data across multiple companies and years in a single execution, which would be challenging with conventional relational databases. Consequently, investors can quickly identify sector-specific performance shifts and evaluate both revenue scale and profitability trends, facilitating more informed strategic decision-making.

\vspace{0.2cm}
\noindent\textbf{Case Study 2: Sector-level stock indicator analysis}

The second case study evaluates sector-wide stock performance and valuation metrics, with a focus on identifying companies with high growth potential or undervaluation within the Semiconductor sector. All companies belonging to this sector were selected using the BELONGS\_TO relationship, and Indicator nodes containing Price-to-Earnings Ratio (PER), Price-to-Book Ratio (PBR), and Earnings Per Share (EPS) for the years 2023, 2024, and 2025 were retrieved. These metrics were aggregated to calculate sector averages and standard deviations, providing a baseline for evaluating relative performance.

The LLM-driven analysis identified companies that are undervalued and characterized by low PER and PBR with consistently positive EPS—such as Lumens (038060), Woori E\&L (153490), LX Semicon (108320), Oditek (080520), and HM Nex (036170). In addition, companies exhibiting strong growth potential, with relatively higher EPS and moderate PER and PBR, were highlighted, including Hanyang Digitech (078350), DB Hitek (000990), IK Semicon (149010), Duksan HiMetal (077360), and Iljin Display (020760). Notably, LX Semicon demonstrated exceptional EPS performance, while Lumens and Oditek were particularly attractive from a valuation perspective due to their low PER and PBR.

This analysis demonstrates that GraphDB supports efficient multi-node aggregation and comparison across companies and years, enabling sector-level insights that would otherwise require complex queries in traditional relational databases. By combining GraphDB with LLM-driven interpretation, investors can identify undervalued stocks and companies with high growth potential within a sector, providing actionable guidance for investment and strategic decision-making. Overall, the study underscores the utility of GraphQA for sector-focused financial analysis, highlighting both valuation and performance trends across multiple years.

%% file: sections/5.Discussion_Conclusion.tex
\section{Discussion \& Conclusion}\label{sec5}

This study presents a novel framework for integrating large language models (LLMs) with a domain-specific Graph Database (GraphDB) constructed using Korean stock market data. The primary contribution of this research lies in three aspects. First, we established a comprehensive GraphDB that consolidates various types of stock-related data, including company information, financial statements, stock prices, and sector classifications. Second, we demonstrated a methodology that leverages LLMs to perform multi-hop reasoning over the GraphDB, enabling the generation of sophisticated and context-aware responses that go beyond simple database queries. Third, through practical case studies, we illustrated the utility of the proposed framework in analyzing real-world stock market scenarios, revealing insights that would be difficult to obtain using conventional data retrieval methods alone.

Despite these contributions, several limitations exist in the current study. The analysis was restricted to the competitive relationships among companies, which limits the exploration of other potentially informative connections, such as supply chain links, shareholder networks, or cross-sector influences. Additionally, the scope of the dataset was confined to the Korean stock market, which may limit the generalizability of the findings to other international markets with different market structures and regulations.

Future research can address these limitations by expanding the scope of the GraphDB to include more diverse types of relationships and by incorporating international stock market data to enable cross-country comparisons. Moreover, enhancing the multi-hop reasoning capabilities of LLMs with additional domain knowledge and temporal dynamics could further improve the depth and accuracy of the insights derived from the database. By overcoming these limitations, the proposed framework has the potential to become a versatile tool for financial analysts, investors, and policymakers seeking to extract actionable intelligence from complex financial datasets.

%% file: sections/6.Appendix.tex
\clearpage  
\section{Property of graph schema}

\begin{table}[h!]
\centering
\begin{tabular}{lp{3cm}p{3cm}}
\toprule
\textbf{Node Type} & \textbf{Property} & \textbf{Description} \\
\midrule
Company  & stock\_code & Stock code \\
         & stock\_nm & Company name  \\
         & stock\_abbrv & Shorted company name \\
         & stock\_nm\_eng & Company name (English) \\
         & listing\_dt & Listing date \\
         & compete\_stock\_nm\_li & List of competitor company names \\
         & compete\_stock\_code\_li & List of competitor stock codes \\
         & market\_nm & Market name \\
         & outstanding\_shares & Number of outstanding shares \\
         & kospi200\_item\_yn & KOSPI 200 inclusion \\
\midrule
Stock    & stck\_oprc    & Open price \\
         & stck\_clpr    & Close price \\
         & stck\_hgpr    & High price \\
         & stck\_lwpr    & Low price \\
\midrule
Indicator & pbr & Price-to-Book Ratio \\
          & per & Price-to-Earnings Ratio \\
          & EPS & Earnings Per Share \\
\midrule
Sector    & stock\_sector\_nm & Standard industry classification name \\
\midrule
Financial & revenue & Revenue \\
Statements& operating\_income & Operating income \\
          & net\_income & Net income \\
          & total\_assets & Total assets \\
          & total\_liabilities & Total liabilities \\
          & total\_equity & Total equity \\
          & capital\_stock & Capital stock \\
\midrule
Date      & date  & Date \\
          & year  & Year \\
          & month & Month \\
          & day   & Day \\
\midrule
Quarter   & year & Year \\
          & quarter & Quarter (numeric) \\
\midrule
Year      & year & Year \\
\bottomrule
\end{tabular}
\caption{Description of properties belonging to nodes in GraphDB}
\label{tab:graphdb_property}
\end{table}

\vspace{5cm}

\section{Case Study}

\vspace{0.2cm}
\noindent\textbf{Case Study 1: Multi-Year Competitive Financial Comparison}

\begin{itemize}[noitemsep, topsep=1pt]
    \item \textbf{Natural Language Question:} 
    "Analyze the performance trends by comparing the revenue, operating income, and net income of Samsung Electronics (stock code: 005930) with its competitor SK Hynix (stock code: 000660) for the years 2023, 2024, and 2025."

    \vspace{0.15cm}
    
    \item \textbf{Question-to-Cypher Conversion Result:} 
    \vspace{-0.35cm}
    

    \begin{listing}[H]
    \caption{Cypher query for case study 1}
    \label{lst:case_study1}
    \begin{scriptsize}
    \begin{lstlisting}[
      frame=single,
      numbers=left,
      basicstyle=\ttfamily\scriptsize,
      columns=fullflexible
    ]
    MATCH (c1:Company {stock_code: "005930"})-[:HAS_FINANCIAL_STATEMENTS]
    ->(fs1:FinancialStatements)-[:FOR_YEAR]->(y:Year)
    MATCH (c2:Company {stock_code: "000660"})-[:HAS_FINANCIAL_STATEMENTS]
    ->(fs2:FinancialStatements)-[:FOR_YEAR]->(y)
    WHERE y.year IN [2023, 2024, 2025]
    RETURN 
      y.year AS year,
      c1.stock_abbrv AS samsung_stock_abbrv,
      fs1.revenue AS samsung_revenue,
      fs1.operating_income AS samsung_operating_income,
      fs1.net_income AS samsung_net_income,
      c2.stock_abbrv AS skhynix_stock_abbrv,
      fs2.revenue AS skhynix_revenue,
      fs2.operating_income AS skhynix_operating_income,
      fs2.net_income AS skhynix_net_income
    ORDER BY y.year ASC
    \end{lstlisting}
    \end{scriptsize}
    \end{listing}
    
    \item \textbf{GraphDB Query Execution Result:} 
    Retrieves multi-year financial metrics—revenue, operating income, and net income—for Samsung Electronics (005930) and its main competitor, SK Hynix (000660), covering the years 2023, 2024, and 2025. The multi-hop query traverses the COMPETES\_WITH relationship to identify competitors and collects corresponding FinancialStatements nodes, enabling a structured dataset for comparative analysis.

    \item \textbf{LLM Answer Generation:} 
    The LLM synthesizes the retrieved data to provide a detailed comparative analysis. It identifies that Samsung Electronics exhibits strong revenue and profit growth in 2024 but experiences a slight decline in 2025, while SK Hynix shows consistent growth across the three years, surpassing Samsung Electronics in operating and net income by 2025. Overall, the LLM highlights that Samsung Electronics maintains a leading revenue scale, whereas SK Hynix demonstrates stronger profitability recovery, offering actionable insights on competitive performance trends in the semiconductor sector.
\end{itemize}

\vspace{1cm}
\noindent\textbf{Case Study 2: Sector-Level Stock Indicator Analysis}

\begin{itemize}[noitemsep, topsep=1pt]
    \item \textbf{Natural Language Question:} 
    "Within the same industry as SK Hynix, identify companies with high growth potential or undervalued stocks based on PER, PBR, and EPS for the years 2023, 2024, and 2025."

    \vspace{0.15cm}
    
    \item \textbf{Question-to-Cypher Conversion Result:} 
    \vspace{-0.35cm}
    \begin{listing}[H]
        \caption{Cypher query for case study 2}
        \label{lst:case_study2}
        \begin{scriptsize}
    \begin{lstlisting}[frame=single, numbers=left, basicstyle=\ttfamily\scriptsize, columns=fullflexible]
    MATCH (sk:Company {stock_abbrv: "SK Hynix"})-[:BELONGS_TO]->(s:Sector)
    <-[:BELONGS_TO]-(c:Company)
    WHERE c.stock_code <> sk.stock_code
    WITH c
    MATCH (c)-[:HAS_INDICATOR]->(ind:Indicator)-[:MEASURED_ON]->(d:Date)
    WHERE d.year IN [2023, 2024, 2025]
    WITH c, ind, d
    WHERE (ind.per < 10 OR ind.pbr < 1 OR ind.eps > 0)
    RETURN 
        c.stock_code AS stock_code, c.stock_abbrv AS stock_abbrv, 
        d.year AS year, ind.per AS per, ind.pbr AS pbr, ind.eps AS eps
    ORDER BY d.year, ind.per ASC, ind.pbr ASC, ind.eps DESC
    \end{lstlisting}
        \end{scriptsize}
    \end{listing}
    
    \item \textbf{GraphDB Query Execution Result:} 
    Retrieves sector-level stock indicators—including Price-to-Earnings Ratio (PER), Price-to-Book Ratio (PBR), and Earnings Per Share (EPS)—for all companies in the Semiconductor sector for the years 2023, 2024, and 2025. The query leverages the BELONGS\_TO relationship to group companies by sector and aggregates indicator nodes to enable sector-wide comparative analysis and identification of outliers.

    \item \textbf{LLM Answer Generation:} 
    The LLM synthesizes the query results to identify companies within the Semiconductor sector that exhibit high growth potential or are undervalued. Undervalued stocks, characterized by low PER and PBR along with positive EPS, include Lumens, Woori E\&L, LX Semicon, O-DITEC, and HM Nex. Companies with strong growth potential, indicated by moderate PER and PBR values combined with increasing EPS trends, include Hanyang Digitech, DB HiTek, IK Semicon, Duksan HiMetal, and Iljin Display. By interpreting PER and PBR as valuation signals and EPS as indicators of profitability and growth, the LLM provides actionable insights to support sector-specific investment decisions.
\end{itemize}